\documentclass[11pt]{article}  
\usepackage{graphicx}
\usepackage[a4paper,left=3cm,right=3.cm, bottom=2.cm, top=2.0cm]{geometry}
\usepackage{amssymb}
\usepackage{amsmath}
\usepackage{multirow}
\usepackage{nicefrac}
\usepackage{csquotes}
\usepackage{bm}
\usepackage[margin=1pt,font=small,labelfont=bf]{caption}
\usepackage[dvipsnames]{xcolor}
\definecolor{citecolor}{RGB}{128,0,32}
\definecolor{headcolor}{RGB}{128,128,128}
\usepackage[unicode,hyperfootnotes=false,breaklinks=true,colorlinks=true,allcolors=citecolor]{hyperref}
\usepackage{setspace}
\usepackage{fancyhdr}
\usepackage{titlesec}
\usepackage{algorithm}
\usepackage{algpseudocode}
\usepackage{etoolbox}

\usepackage{libertinus}
\usepackage{tabularx}
\usepackage{datetime} 
\usepackage{etoolbox}
\usepackage{subcaption} 
\patchcmd{\linenumber}{\hb@xt@}{\hbox}{}{}
\usepackage[ giveninits=true, style=authoryear, backend=biber, maxcitenames=2, maxbibnames=99,
  hyperref=true, backref=false, sorting=nyt, uniquename=init, autolang=hyphen,
  dashed=false]{biblatex}

\AtEveryCite{\it \color{citecolor}} \AtEveryBibitem{\clearfield{month}}
\AtEveryBibitem{\clearfield{issn}}
\renewbibmacro{in:}{}

\let\citeA\textcite
\let\cite\parencite
\graphicspath{{./newfigs/}}
\newdateformat{usvardate}{\shortmonthname[\THEMONTH]. \THEDAY, \THEYEAR}
\newcommand{\lastmodified}{%
  \textcolor{headcolor}{\scriptsize \usvardate\today{} at \currenttime} }
\fancypagestyle{custom}
{      
\fancyhead[L]{}
\fancyhead[C]{\textcolor{headcolor}{\scriptsize submitted to \textit{Geochemistry, Geophysics, Geosystems}}}
\fancyhead[R]{}
\fancyfoot[C]{\textcolor{gray}{\thepage}}
\fancyfoot[R]{\lastmodified}
\fancyfoot[L]{}
}
 \titlespacing\section{0pt}{12pt plus 4pt minus 2pt}{0pt
plus 2pt minus 2pt}

\begin{document}

\pagestyle{custom}
\begin{center}
\LARGE {\bf 3-D numerical modelling of the feedback between deformation and thermal structure during subduction initiation for the French Lesser Antilles\\[12pt]}

\normalsize
Ekeabino Momoh$^{1,2,\P}$,
Stephen Tait$^{1,2}$,
Harsha S. Bhat$^{3}$
\\[12pt]

\begin{enumerate}
	\scriptsize
	\setlength\itemsep{-5pt}
  	\item {Geosciences Environnement Toulouse - GET, Universit\'{e} Toulouse III - Paul Sabatier,  CNRS UMR 5563, {31400} Toulouse, France}
    \item {Institut de Physique du Globe de Paris, UMR 7154 Universite de Paris {75005} Paris, France}
    \item {Laboratoire de Géologie, École Normale Supérieure, CNRS UMR 8538, PSL Research University, 75005 Paris, France}    
\end{enumerate}

\let\thefootnote\relax\footnotetext{$\P$ Corresponding author: \texttt{ekeabino.momoh@get.omp.eu}}
\end{center}

\section*{Key Points}
\begin{itemize}
  \small
  \item 3-D insights on pre-subduction initiation and heat production from inelastic deformation in an intra-oceanic domain with crustal offsets.
  \item Contribution to  thermal structure and location of partial melting from volumetric and deviatoric strain localization. 
  \item Establishment of a thermal steady-state.
\end{itemize}

\section*{Abstract}
\small
We used 3-D thermomechanical modelling to investigate conditions during subduction-zone initiation and early thermal development with focus on the Lesser Antilles. Our model imposes a convergence velocity of 2 cm per year and incorporates heating caused by irreversible deformation of mantle and crustal rocks, using elasticity, creep, and non-associative plastic flow laws. Our results show that deformational heating before slab development is unexpectedly strong. After several million years, buckling and heating due to irreversible deformation create distinctive patterns of topography and surface heat flow that resemble present-day observations, despite the slab and subduction interface being incompletely developed. Within the Caribbean plate, plate buckling produces a high topographic ridge underlain by a large positive thermal anomaly of approximately 200 K, centred just below the Moho. The conductive thermal boundary layer transporting this heat to the surface thins from about 100 km to 10 km beneath the topographic maximum, allowing the ridge to rise above sea level. This thermal structure suggests the potential initiation of a volcanic arc approximately 180 km from the inter-plate contact. A hot zone at 30-50 km depth has pressures consistent with those inferred from Lesser Antilles primitive magmas and represents the most plausible location for partial melting of Caribbean mantle if volatiles are present. The thick Caribbean crust, approximately 20-25 km, is also heated sufficiently for possible silicic melt generation. The inferred lithospheric thickness of 50-100 km aligns with tomography studies. Thus, subduction thermal structure is strongly influenced by several million years of initiation processes.

\section*{Plain Language Summary}
The process of subduction initiation has been an enigmatic subject for decades. Here, we explore how the thermal state is strongly related to the stress and strain fields due to deformational heating, that develop during initiation, rather than mantle upwelling driven by corner flow, i.e., the plausible steady state. The main lithospheric structures in the overriding plate are bands exhibiting both shear and volumetric (dilatational) plasticity during irreversible plastic flow, which delimit the high topography of the arc. These bands extend via ductile creep into the underlying asthenosphere. Dilatational plasticity and its energetic role have been mostly neglected historically in geodynamics. Our results are consistent with the observed distance of the volcanic arc from the interplate contact and the published heat flow measurements. Our results suggest that deformational heating can lead to mafic partial melt generation in the mantle at depths of $\sim$40-50 km, as well as silicic melt in the lower Caribbean crust, after only a few million years at the specified convergence rate of 2 cm per year if the Caribbean mantle and crustal solidi are lowered enough by the presence of volatiles. 

\section{Introduction and Geological Underpinning}
The Lesser Antilles is an 850 km volcanic island chain in the Atlantic, stretching from Sombrero, Anguilla (Virgin Islands) in the North to Grenada in the South \cite{davis1924formation,wadge1994lesser}. It formed following the convergence of the Atlantic plate beneath the Caribbean plate at a low rate of 2 cm/year, with relative plate motion constant over the last 38 million years, with a radius of curvature of 450 km \cite{westbrook1984geophysics,maury1991geology,macdonald2000lesser}. Over the last 100 thousand years, it has been characterised by a low magma flux \cite{wadge1984comparison,macdonald2000lesser}. At present, the locus of magmatism extends from Grenada to Saba, with ages traced to Eocene (${\sim}$56 million years) in Martinique and as recent as 2.3 million years ago further North within the arc \cite{bouysse19904,wadge1994lesser}. Crustal thicknesses were estimated to be 15 to 30 km beneath the forearc and volcanic arc areas, and less than 10 km for the subducting (Atlantic) plate, composed of more typical oceanic crust \cite{boynton1979seismic,westbrook1984geophysics,kopp2011deep,klingelhoefer2025lateral}.  Currently, the Caribbean plate is interpreted as an 8 km- to 20 km-thick oceanic plateau, formed atop the Galapagos hotspot, before collision with the pristine Caribbean  arc and Northwestern segment of South America \cite{mauffret1997seismic,buchs2016evidence,yang2022natural}. {{Recent geophysical investigations targeted the LAA as a case of subduction of particularly cold and volatile-rich oceanic lithosphere \cite{goes2019project}. The resultant tomographic images clearly show along-arc variations in a number of observables, possibly related to amounts of volatiles, melts and/or thermal structure \cite{hicks2023slab,bie2022imaging} which challenge existing 2-D geodynamic models \cite{perrin2016reconciling}. The LAA volcanoes are known to be embedded in an array of active faults with mostly normal and strike-slip source mechanisms, representing intraplate (Caribbean) deformation, whose main role has been to allow arc parallel extension \cite{feuillet2002arc}. The above observations emphasise the need for a 3D approach to geodynamic modelling of this plate convergence, and raise the question of the spatial position of the volcanoes with respect to the deformation.}}

Subduction (convergent) zones, where mantle lithosphere descends into the Earth's interior like the Lesser Antilles, represent some of the most seismically and volcanically active tectonic regions in the globe. The formation of a subduction zone involves a stage of strain localisation in a compressional tectonic regime during which major plate-boundary faults develop \cite{regenauer1998rapid}, followed by a stage of thermal evolution towards a steady-state. The processes involved in the formation/development of a new subduction zone have not been well understood, nor is the thermal structure explicitly known. A consensus view in the literature is that one indispensable ingredient in the physical description of steady-state subduction is a so-called corner flow, whereby the downgoing lithospheric plate with prescribed velocity viscously entrains mantle wedge material \cite{mckenzie1969speculations}. Numerous authors have indeed explored the thermal structure that becomes established as heat removal by conduction comes to balance that advected in by the corner flow \cite{van2003structure,syracuse2010global,england2010melting}. Nevertheless, it has been recently pointed out that the surface heat flux in the ensuing model steady state is much lower than that typically observed at the arc \cite{jones2018thermal} implying a need to re-examine the energy budget carefully. So while we do not deny the likelihood of a corner flow once subduction is fully developed, it can only be established after a protracted initial period of deformation. It is worth asking what the potential influences during this stage of initiation could be. 

\begin{figure}
\centering{\includegraphics[width=15cm]{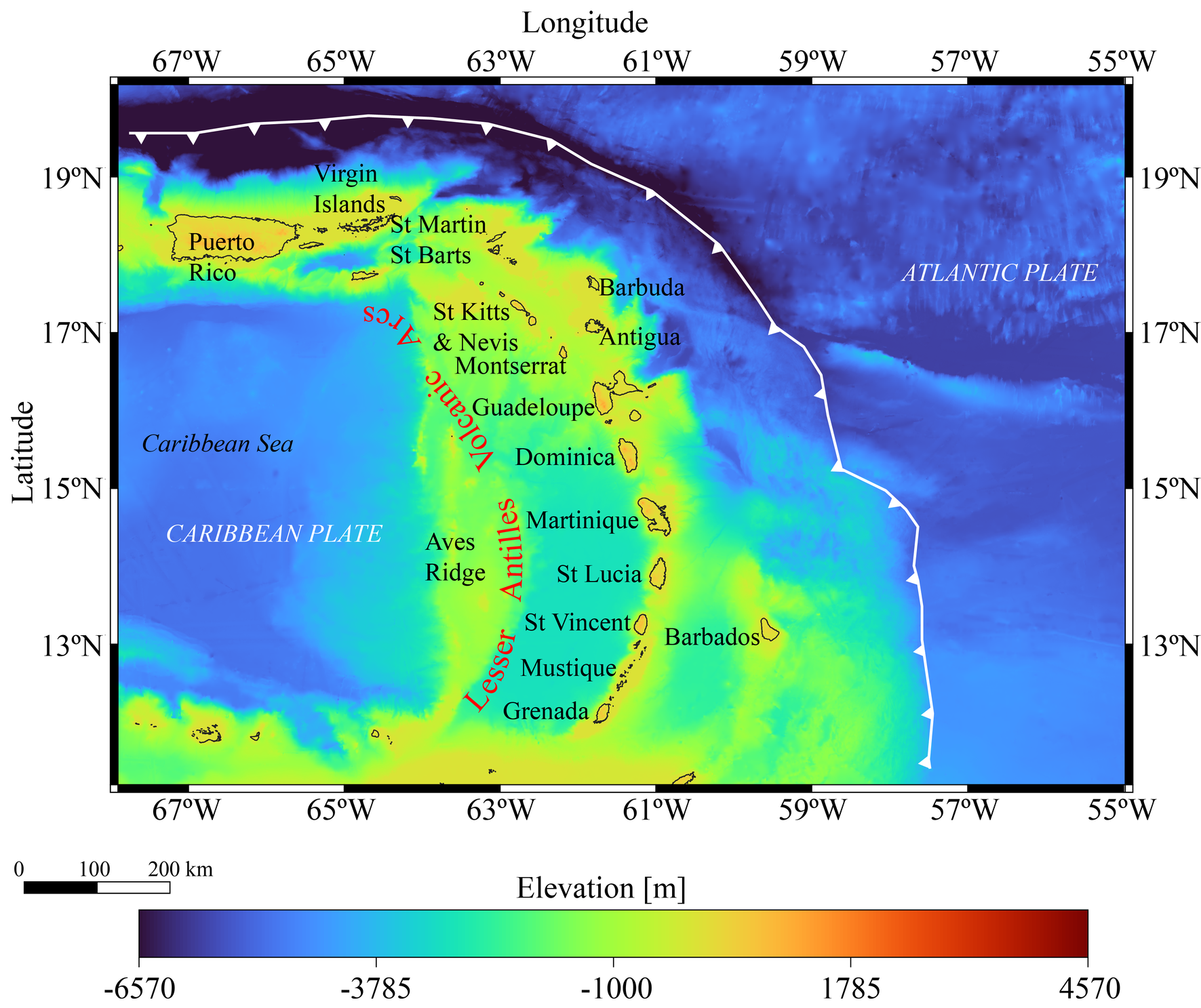}}
\caption{Topographic map for the Lesser Antilles area showing the volcanic archipelago and the subduction trench (in white). We note that individual volcanic edifices are small with respect to the size of the islands that host them, which are in turn  much smaller than the broad topographic high of which they are just the emergent part. The data for the map was obtained from the global terrain model of General Bathymetric Chart of the Oceans data was obtained from \url{https://www.gebco.net/data-products/gridded-bathymetry-data}. The map shown here has been created using the Free and Open Source QGIS \cite{QGISsoftware}.}
\label{momoh_map}
\end{figure}

In a bid to analyse this sequence of events and improve understanding of the deformation prior to steady subduction, we report on model simulations which (within the bounds of the chosen rheologies) both explicitly follow the strain localisation process and the thermal evolution of the deformed zone and surrounding regions. We include specifically the heat source term that is due to irreversible deformational work in plastic and ductile creep regimes, and incorporate a rheologic description appropriate for geologic materials of the upper mantle and oceanic crust. The distribution of the heating effect is linked to the
deformation field at the scale of the deformation bands that develop and to how their orientations cause them to interact. Localised heating can both lead to further localisation or to large hot zones where strong shear occurs in competing directions. The current paper reports on 3-D geodynamic simulations incorporating these complex feed-back loops at larger (geological) scale with a specific target at a segment of the French Lesser Antilles in the context of international, national and regionally funded science efforts and the permanent monitoring activities of the French Volcanic and Seismic Observatories. The regional bathymetric map (Figure \ref{momoh_map}) shows how the volcanic arc is built on a broad topographic high: for example, the breadth of the region with ocean depth less than 1000 m is on the order of 200 km wide. This is much greater than the scale of the individual volcanic edifices or even the islands. The Aves ridge is another intriguing bathymmetric feature whose origin is poorly understood. Our prime focus will be the Lesser Antilles volcanic arc.

\section{Methodology}
We utilised a thermomechanical constitutive law for solid deformation which implements elastic, creep and viscoplastic rheologies. We modelled dislocation creep as the dominant creep mechanism, which is activated for the entire model, but significant only where the temperature is high enough for creep deformation at the local strain rate. Viscoplastic flow is activated when a pressure-dependent Drucker-Prager yield criterion is overcome by the stress state. Details of this algorithmic implementation have been elaborated in \citeA{momoh2025volumetric}. The basic elements include small strain theory, solid mechanics, a non-linear temperature-dependent creep correction, and a rate-dependent Drucker-Prager plasticity, including volumetric plastic strains. For plastic deformation, we have also included friction hardening and a linear cohesion hardening, which leads to a kinematic evolution of the yield surface. For elasticity, we have accounted for thermoelastic effects. We provide the elements in Table \ref{conservation_and_constitutive_laws}. 

\begin{table}[p!]
\centering
\small
\caption{Conservation and constitutive laws for deformation and stress update. }
\label{conservation_and_constitutive_laws}
\begin{tabular}{  m{0.5cm}  m{6cm} m{7cm}  } 
  \hline
\multicolumn{3}{l}{\textbf{1. Conservation laws}}   \\[2pt] 
\multirow{2}{4em} & 
(a){$\dfrac{\partial{{\sigma}_{ij}}}{\partial{x}_j} +f_i=\rho\dfrac{\partial^2{u_i}}{\partial{t}^2}$}
\vskip 10pt
(b) ${\dfrac{\partial{T}}{\partial{t}}=\underbrace{\alpha_\textrm{th}{\nabla^2 T}}_\mathrm{diffusivity}+\underbrace{\beta\dfrac{{\sigma_{ij}}\left(\dot\varepsilon^\textrm{v}_{ij}+\dot\varepsilon^\textrm{vp}_{ij}\right)}{\rho C_p}}_\mathrm{deformational\;heating}}$\newline${\;\;\;\;\;\;\;\;\;\;\;\;\;\;\;\;\;\; -\underbrace{\dfrac{C_L}{C_p}\dfrac{\partial}{\partial t}F(p,T)}_\mathrm{latent\;heat}}$ 
& (a) ${\sigma_{ij}}$ are stress tensor components, 
${f_i}={\rho g}$ are body forces,
${\rho}$ is density, ${g}$ as gravitational acceleration, ${x_j}$  represents spatial variables in the Cartesian coordinates; ${{u_i}}$ are the components of displacement. \vskip 10pt (b) ${T}$ is temperature, ${{C_p\;\mathrm{(J\;kg^{-1}\;K^{-1})}}}$ is specific heat capacity and ${\alpha_\textrm{th}\;\mathrm{(m^{2}s^{-1})}}$ is the thermal diffusivity. ${{{\dot\varepsilon}}^{\textrm{v}}_{ij}}$ and ${{{\dot\varepsilon}}^{\textrm{vp}}_{ij}}$ are creep and viscoplastic strain rate components, respectively ${\beta}$, is the fraction of deformational work converted to heat. ${{C_L\;\mathrm{(kJ\;kg^{-1})}}}$ is the latent heat where ${F (p,T)}$ represents the melt fraction. \\ 
\hline
\multicolumn{3}{l}{\textbf{2. Strain (rate) decomposition}}   \\[0.5pt] 
 & (a) ${{\dot\varepsilon}_{ij} = {\dot\varepsilon}^\textrm{e}_{ij}+{\dot\varepsilon}^\textrm{v}_{ij}+{\dot\varepsilon}^\textrm{vp}_{ij}}$\newline (b) ${{\dot\varepsilon}_{ij} = {\dot\varepsilon}^\textrm{e}_{ij}+{\dot\gamma^\textrm{v}}\dfrac{\partial\Phi^\textrm{v}_\textrm{F}}{\partial s_{ij}}+{\dot\gamma^\textrm{vp}}\dfrac{\partial\Phi^\textrm{vp}_\textrm{F}}{\partial\sigma_{ij}}}$ \newline &(a) \enquote{$\textrm{e}$}, \enquote{$\textrm{v}$} and \enquote{${\textrm{vp}}$} represent elastic, creep and viscoplastic, respectively.\newline \newline (b) ${\Phi^\textrm{v}_\textrm{F}}$ and ${\Phi^\textrm{vp}_\textrm{F}}$ are creep and viscoplastic flow potentials, respectively, and ${{\dot\gamma}^\textrm{v}}$ and ${{\dot\gamma}^\textrm{vp}}$ are creep and viscoplastic multipliers in rate form; ${s_{ij}}$ are deviatoric stress tensors; ${{s_{ij}=\sigma_{ij}}-\sigma_{kk}/3}$.  $J_\textrm{II}=s_{ij}s_{ij}/2$ is the deviatoric stress invariant.\\ 
 \hline
\multicolumn{3}{l}{\textbf{3. Elastic trial state and creep updates}}   \\[2pt]  
  \multirow{2}{4em} & (a) ${{\sigma}_{ij}^\textrm{e}=C_{ijkl}^\textrm{e}({\varepsilon}_{kl}^\textrm{e}-\alpha\Delta T\delta_{kl})}$ \newline(b) ${\dot\varepsilon_{ij}^\textrm{v}=\dot\gamma^\textrm{v}\dfrac{s_{ij}^\textrm{e}}{\sqrt{J_\textrm{II}\left({s_{ij}^\textrm{e}}\right)}}}$ \newline (c) ${s_{ij}^\textrm{v}=\left(1-\dfrac{G\Delta\gamma^\textrm{v}}{\sqrt{J_\textrm{II}}}\right)s_{ij}^\textrm{e}}$ & (a) Elastic stress state. ${C_{ijkl}}$ is the elastic stiffness matrix, ${\alpha}$ is thermal expansivity, ${\varepsilon_{kl}^\mathrm{e}}$ are components of elastic strain, .\newline\newline  (b) Creep strain rate update \newline\newline (c) Deviatoric stress update due to creep deformation.\\ 
\hline
\multicolumn{3}{l}{\textbf{4. Plastic state and updates}}\\
 \multirow{3}{4em} &(a) ${\Phi_{Y}({\sigma}_{ij}, c) = \sqrt{J_\textrm{II}^\textrm{v}}+\alpha_1P-\alpha_2c \ge 0}$\newline (b) ${{\dot\varepsilon}^\textrm{vp}_{ij}  = {\dot\gamma^\textrm{vp}}\left(\underbrace{\dfrac{{s}^{\textrm{v}}_{ij}}{2\sqrt{J_\textrm{II}^\textrm{v}({s}^{\textrm{v}}_{ij})}}}_\textrm{deviatoric}+ \underbrace{\dfrac{\alpha_3}{3}\delta_{ij}}_\textrm{volumetric}\right)}$ \newline (c)${{{{s}_{ij}^\textrm{vp} ={\left(1-\dfrac{G\Delta{\gamma^\textrm{vp}}}{\sqrt{J_\textrm{II}^\textrm{v}({s}^{\textrm{v}}_{ij})}}\right)}{s}_{ij}^{\textrm{v}}}}}$ \newline (d)${P=P^{\textrm{e}}-\underbrace{{\alpha_3}\Delta{\gamma^\textrm{vp}}K}_\textrm{volumetric}}$& (a) Drucker-Prager yield criterion \newline\newline(b) Viscoplastic strain update\newline\newline(c) Deviatoric stress update\newline\newline(d) Pressure update. ${K}$ represents bulk modulus, ${\alpha_3}$ depends on plastic dilatancy.\\
\hline
\multicolumn{3}{l}{\textbf{5. Model for partial melt}}\\
\multirow{2}{4em}&${F(p,T) = \left(\dfrac{T-T_{\textrm{solidus}}}{T_{\textrm{liquidus}}-T_{\textrm{solidus}}}\right)^{1.5}}$&${\textrm{for} \; T_{\textrm{solidus}}< T < T_{\textrm{liquidus}}}$,\newline ${F(p,T)}$ represents the degree of melt as a function of temperature.\newline ${T_{\textrm{solidus}} =A_1P + A_2P + A_3P^2}$\newline${ 
T_{\textrm{liquidus}} =B_1P + B_2P + B_3P^2}$	\\
\hline
\end{tabular}
\end{table}
Plastic yielding is described by:
\begin{equation}
{\Phi_{Y}({\sigma}_{ij}, c) = \sqrt{J_\textrm{II}^\textrm{v}}+\alpha_1P-\alpha_2c \ge 0},
\end{equation}
where ${J_\mathrm{II}^\mathrm{v}}$ is the updated viscoelastic deviatoric stress invariant, ${P}$ is the pressure, and ${c}$ is the cohesion which may depend on the viscoplastic strain history or can be constant; ${\alpha_1}$ and ${\alpha_2}$ which are used in defining the yield state are material-dependent constants which are given functions of the internal friction angle (${\varphi}$) as follows:
\begin{equation} 
\alpha_1=\dfrac{6 \sin{\varphi}}{\sqrt{3}(3-\sin\varphi)}, \alpha_2=\dfrac{6\cos{\varphi}}{\sqrt{3}(3-\sin\varphi)}.
\end{equation}
We also consider expansion during plastic flow (dilatancy) described by a material parameter (${\alpha_3}$), the dilatancy angle. We also incorporated two hardening laws in
our constitutive formulation. First is a linear cohesion hardening
which depends on the deformation history and second is friction
hardening. The cohesion hardening is given by:
\begin{equation}
c=c_0+H\bar\varepsilon
\end{equation}
where ${c}$ is the updated cohesion, ${c_0}$ is the initial cohesion, ${H}$ is the hardening modulus, and ${\bar\varepsilon}$ is a quantity that records the deformation history during viscoplastic deformation, which can be seen as accumulated viscoplastic strain: ${\dot{\bar\varepsilon}}=\alpha_2\dot{\gamma}$, or in incremental form, ${\Delta{\bar\varepsilon}}=\alpha_2\Delta{\gamma}$. The hardening modulus can be interpreted as the slope of the stress-strain curve during plasticity, i.e., an incremental yield surface as a function of the viscoplastic deformation history.
\begin{equation}
c(\bar{\varepsilon}^\textrm{vp})=  c_0 + H\bar{\varepsilon}_{n+1}^\textrm{vp} = c_0 + H(\bar{\varepsilon}_{n}^\textrm{vp} + \Delta{\bar{\varepsilon}^\textrm{vp}})= c_0 + H(\bar{\varepsilon}_{n}^\textrm{vp}+\alpha_2\Delta{\gamma}).
\end{equation}
For frictional hardening, we use the form of \cite{leroy1989finite,leroy1990finite} for monotonically increasing the friction angle with the deformation:
\begin{equation}
sin\;\varphi=sin\;\varphi_\textrm{i}+\dfrac{2\left(sin\; \varphi_\textrm{f}-sin\;\varphi_\textrm{i}\right)\sqrt{\varepsilon_\textrm{eff}^\textrm{vp}\varepsilon_\textrm{crit}^\textrm{vp}}}{\varepsilon_\textrm{eff}^\textrm{vp}+\varepsilon_\textrm{crit}^\textrm{vp}}.
\end{equation}
Where the effective friction angle ${\varphi}$ increases from an initial friction angle ${\varphi_\textrm{i}}$ to a final friction angle ${\varphi_\textrm{f}}$, where the final friction angle is attained at a critical strain  ${\varphi_\textrm{crit}}$. The effective viscoplastic strain which is a measure of the accumulated viscoplastic deformation is given by:
\begin{equation}
\varepsilon_\textrm{eff}^\textrm{vp}=\sqrt{\dfrac{2}{3}\varepsilon_{ij}^\textrm{vp}\varepsilon_{ij}^\textrm{vp}}.
\end{equation}

As a simple initial approach, we considered a melting model for anhydrous peridotite which follows the power-law melting formulation of \cite{hirschmann2000mantle,katz2003new}, given in Table \ref{conservation_and_constitutive_laws}. ${F(p,T)}$ represents the degree of melt as a function of temperature (in ${{\textrm{T}}}$) and pressure (in ${\textrm{GPa}}$). In the subduction context it has been much more common to consider constitutive models for melting that involve volatiles, based on the idea that dehydration reactions in the slab will supply volatiles \cite{van2003structure,syracuse2010global}. Although this is expected to be an important phenomenon at steady state, it should not be a major point during subduction initiation before a slab has reached depths on the order of 100 km. Hence, our use of an anhydrous model at this stage. We also do not include a description of melt transport by permeable flow, as our main focus is to describe the heating up of parts of the domain to see where, if anywhere the temperature comes to exceed the local solidus. This amounts in common petrologic parlance to assuming batch melting, i.e., such that mantle rocks melt without an instantaneous melt extraction, which therefore does not take account of the melting history in our model or requires a modification of the description of the melted units. The constants used were: ${A_1}$,${A_2}$,and ${A_3}$ = 1085.7${^\circ}$C GPa\textsuperscript{-1}, 132.9${^\circ}$C GPa\textsuperscript{-1} and -5.1${^\circ}$C GPa\textsuperscript{-1}, respectively; and ${B_1}$, ${B_2}$, and ${B_3}$ = 1475${^\circ}$C GPa\textsuperscript{-1}, 80${^\circ}$C GPa\textsuperscript{-1} and -3.2${^\circ}$C GPa\textsuperscript{-1} \cite{hirschmann2000mantle,katz2003new}. 

\noindent Results shown are:

\begin{equation}
\sqrt{J_\mathrm{II}}= \sqrt{\dfrac{1}{2}s_{ij}^\mathrm{vp}s_{ij}^\mathrm{vp}},\; {\varepsilon_\mathrm{II}^\mathrm{IN}}= \sqrt{\dfrac{1}{2}\varepsilon_{ij}^\mathrm{vp}\varepsilon_{ij}^\mathrm{vp}},\;\varepsilon_\mathrm{v}^\mathrm{IN}=\varepsilon_{kk},\; \mathrm{and}\; \mathrm{log_{10}}\dot\varepsilon_{ij}=\mathrm{log_{10}}\left(\sqrt{\dfrac{1}{2}\dot\varepsilon_{ij}\dot\varepsilon_{ij}}\right);
\end{equation}
representing the second invariant of deviatoric stress (an invariant measure of shear stress), second invariant of inelastic deviatoric strain (magnitude of deviatoric deformation), first invariant of strain (magnitude of volumetric deformation), and logarithm of the second invariant of the deviatoric strain rate tensor, incorporating elastic, creep and plastic contributions. The superscript (\enquote{IN}) represents inelastic, i.e., quantities are plotted after creep and plastic correction. The volumetric changes in deformation are driven by dilatant plasticity. Additional plots shown are temperature and melt fraction. The temperature is updated due to inelastic work while the melt fraction is checked after every temperature change. 

The key takeaway from this technical section is that we implement a full constitutive update at each time step to ensure that the parts of the model responding in an elasto-plastic way (i.e., the lithosphere) and those responding via ductile creep (i.e., the asthenosphere) are not fixed but continuously followed throughout a given simulation. We thus account for temperature, stress and rate dependency of the local rheologic response during deformation. Ductile rocks which define the asthenosphere are identified where the normal stresses become equal, and thus approximate to an incompressible fluid. In the lithosphere, by contrast, normal stresses are not equal (their average can also differ from $\rho g y$)  and hence some volumetric plastic deformation occurs. The presence of volumetric strain thus gives us one proxy to visualise the lithosphere, which would be susceptible to tectonic seismicity. Observed tectonic seismicity provides one dataset related to lithospheric thickness to compare with our output of strain.

\section{Numerical Simulations and Results}
\label{modelling_and_results}
\subsection{Model Setup and Boundary Conditions}
\begin{figure}
\centering{\includegraphics[width=12cm]{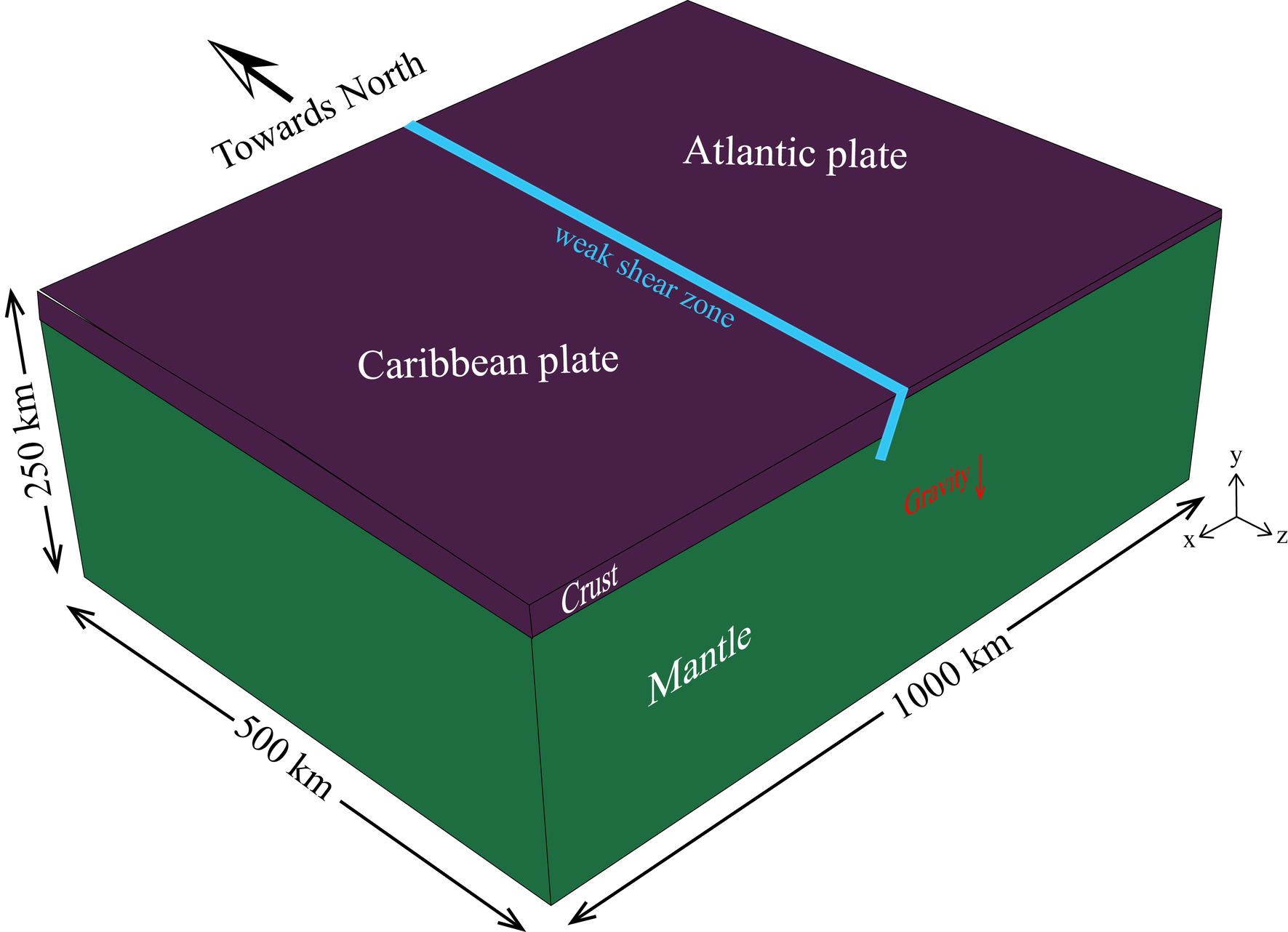}}
\caption{Input 3-D model with Atlantic plate on the East encompassing an 8 km-thick oceanic crust and the Caribbean plate hosting a 20 km oceanic plateau with the crustal offset co-located with a plastically and elastically weak deformation zone. A 2 cm/year convergence velocity is imposed from the Eastern boundary, while inward normal velocities have been set to zero for the other vertical and bottom boundaries, i.e., they are free to slip in tangential directions. The top surface is left free to deform.}
\label{momoh_input_model}
\end{figure}

The constitutive updates are implemented in the Abaqus finite element solver \cite{abaqus19}. For present purposes, the main impact of the geological history on our model setup is that we introduce a contrast in crustal thickness at the prescribed locus of subduction initiation, although only one generic oceanic crust rheology is employed for the Caribbean and Atlantic plates. We do not include oblique convergence yet. The subducting Atlantic plate is composed of an 8 km oceanic crust, while the Caribbean plate is composed of a 20 km oceanic plateau (Figure \ref{momoh_input_model}). These crustal layers are underlain by a homogeneous mantle. An 80-million-year-old oceanic lithosphere was used to describe the initial geotherm, while the bottom and top boundaries have constant temperatures, with zero heat fluxes on all vertical boundaries. Rheologic parameters for mantle and crustal rocks are given in Table \ref{matpros}. We impose a velocity of 2 cm/yr on the eastern edge of the model (held constant) to kinematically induce convergence of these two plates, while normal velocities are left at zero on all other vertical boundaries. While these assumptions are at least partly based on what is observed today, as the conditions at inception are unknown, our assumptions are intended to be as simple as possible. The convergence does appear to have been approximately steady (ref) during $\sim{40}$ myrs \cite{lallemand2021subduction}. Our main goal was to assess how transient initiation processes might influence the structure at the onset of steady subduction. Subduction initiation requires a weakness/suture zone, which can be accomplished by grain size reduction, rheological offsets between a weak zone and the adjacent areas, thermal heterogeneity, age-offsets, density differences, and anisotropy. In the calculation shown here, we demarcated the crustal offset with an elastically and plastically weak suture zone. The main role of the weak zone is to initiate the localisation of deformation. In presenting our results, we first consider the possible timescale on which the system may evolve (Section \ref{timescaleevolution}), then describe transient evolution of strain and temperature (Section \ref{stress_matching}), quasi-steady thermal structure and associated deformational state (Section \ref{quasisteadystate}), then weigh implications for partial melting and volcanic arc location (Section \ref{partialmeltingimplications}). 

{{\begin{table}\begin{center}
\begin{minipage}{113mm}
\caption{Material properties in the geodynamic model. \hspace{2mm}${E}$ (Young's modulus), $v $ (Poisson's ratio), $K$ (bulk modulus), $\rho_\mathrm{ref}$  (reference density), $g$ (gravitational constant), ${C_p}$ (specific heat capacity), ${C_p}$ (latent heat), $\alpha_\textrm{th}$  (thermal diffusivity), $\alpha_\textrm{ex}$  (thermal expansivity coefficient), ${m}$ stress sensiticity exponent during creep, ${A}$ (pre-exponential multiplier), $E_a$  (creep activation energy), $R$ (molecular gas constant), ${\varphi_i}$ (initial friction angle), ${\varphi_f}$ (final friction angle), ${\psi}$ (dilatancy angle), $c_0$ (initial cohesion), $H$  (hardening modulus), ${1/\mu}$ (relative rate of viscoplastic strain), ${\varepsilon_\mathrm{crit}}$ is the critical accumulated plastic strain at which the friction angle reaches its final value and is maintained constant, and $m$ (stress exponent during viscoplastic flow). Thermal conductivity was 2.25 ${\mathrm{W^{-1}\;m^{-1}\;K^{-1}}}$. The material parameters for dislocation creep and thermal properties were drawn from \cite{currie2015geodynamic,babeyko2008high}. Values for mechanical parameters were drawn from \cite{turcotte2002geodynamics}.  Plasticity parameters were drawn from \cite{babeyko2008high}.}
\label{matpros}
\begin{tabular}{@{}lllllll}
\hline
Material property  (unit):  &  Mantle & Oceanic crust & Weak zone\\
\hline
Mechanical &           &      &  \\[2pt]
\hspace{2mm}${E}$ (GPa) & 140     & 60 & 100\\[2pt]
\hspace{2mm}$v$    & 0.25     & 0.25 & 0.25 \\[2pt]
\hspace{2mm}$\rho_\mathrm{ref}$ (kg m${^\textrm{-3}}$)    & 3300     & 2950 & 3200\\[2pt]
\hspace{2mm}$ g $ (m s${^{\textrm{-2}}}$)       & 9.8     & 9.8& 9.8
\\[2pt]
\hline
Thermal &            &      \\[2pt]
\hspace{2mm}${C_p}$  (J kg${^\textrm{-1}}$K${^\textrm{-1}}$)       & 1250& 750 & 1250\\[2pt]
\hspace{2mm}${C_L}$  (kJ kg${^\textrm{-1}}$)       & 450& 450 & 450\\[2pt]
\hspace{2mm}$\alpha_\textrm{ex} {\;}$ (K${^\textrm{-1}}$) &  ${3.5\times{10^{-5}}}$     &  ${3.5\times{10^{-5}}}$ &  ${3.5\times{10^{-5}}}$\\
\hline
Dislocation creep     &         &  \\[2pt]
\hspace{2mm}$m $  & 3     & 4.7    & 4.7 \\[2pt]
\hspace{2mm}${A}$ (Pa${^\textrm{-m}}$ s${^\textrm{-1})}$       & $3.91\times10^{-15}$   &$5.05\times10^{-28}$&$5.05\times10^{-28}$\\[2pt]
\hspace{2mm}$E_a\; $ (kJ mol${^\textrm{-1})}$    & 532 & 430 & 485 \\[2pt]
\hspace{2mm}$R\;$ (J K${^\textrm{-1}}$mol${^\textrm{{-1}}}$)  & 8.31     & 8.31 & 8.31 \\
\hline
Plasticity &     &        &      &  \\[2pt]
\hspace{2mm}${\varphi_i {\;} (^\textrm{o})}$&  28     & 20 & 15\\
\hspace{2mm}${\varphi_i {\;} (^\textrm{o})}$&  36.9     & 25 & 20\\
\hspace{2mm}${\psi {\;} (^\textrm{o})}$      & 15     & 10 & 10\\[2pt]
\hspace{2mm}$c_0 {\;} $(MPa)    & 10         & 10 & 1 \\[2pt]
\hspace{2mm}$H {\;} $ (GPa)           & 1     & 1 & 0.5\\[2pt]
\hspace{2mm}${1/\mu}$ (s${^ {-1}}$)    & ${10^{-15}}$       & ${10^{-15}}$    & ${10^{-15}}$ \\[2pt]
\hspace{2mm}${\varepsilon_\mathrm{crit}}$ (-)    & 0.005       & 0.005    & 0.005 \\[2pt]
\hspace{2mm}$m $  & 1     & 1    & 1 \\[2pt]
\hline
\end{tabular}\\
\end{minipage}
\end{center}
\end{table}}}

\subsection{Predicted timescale of thermal evolution}
\label{timescaleevolution}
The energy balance equation contains two timescales which derive from the deformational heating rate and the thermal conduction terms, respectively. The timescale of deformation is derived from the far-field boundary velocity and the length scale over which the rheology varies from elasto-plastic to ductile (mainly because of the geothermal gradient). This latter length is a (thermal) measure of the lithospheric thickness (say $L_T$) which is an unknown a priori. The deformation timescale, if we assume say an initial $L_T$ of $\sim100 $km, is thus $\tau_\mathrm{def}\approx L_T/v_0 \sim10^{14}$s, i.e., $ \sim$3 Myrs. The conductive timescale, written in terms of $L_T$ i.e., $ \tau_\mathrm{con}\approx L_T^2/\alpha_\textrm{th}\approx  10^{16}$s, is $ \sim$300 Myrs. This tells us that $\tau_\mathrm{def}$ is 100 times shorter than $\tau_\mathrm{con}$ and hence we expect that the initial evolution will be on the timescale of $\tau_\mathrm{def}$, i.e., a few million years. Conduction is initially ineffective at evacuating heat from a deep source to the surface. Therefore, we first illustrate the evolution of the system (via strain and stress invariants as well as temperature increase) as a series of snapshots of our results that cover the first $\sim$ 7 Myrs (Figure \ref{time_steps_evolution_of_EII_T}). By this stage, the system has absorbed $\approx$140 km of convergence. 

The system can attain thermal steady state during deformation if the conduction term and the heat production term come into approximate balance, i.e, if the conductive timescales and deformational timescales become similar, which is possible if $L_T$ decreases with respect to the above preliminary estimate. It is not obvious that stress and strain rates will stabilise to roughly spatially steady values. To examine this possibility more closely, we write the steady state version of the energy equation shown in Table \ref{conservation_and_constitutive_laws}.

\begin{equation}
 0 \approx {\alpha_\textrm{th}{\nabla^2 T}} +{\beta\dfrac{{\sigma_{ij}}\left(\dot\varepsilon^\textrm{v}_{ij}+\dot\varepsilon^\textrm{vp}_{ij}\right)}{\rho C_p}}. 
\end{equation}

We now non-dimensionalize this equation using the following scales: temperature (${T_S - T_0}$), length (${L_T}$), time (${L_T/v_0}$), and stress (${\rho g L_T}$). We thus scale temperature with the difference between the solidus temperature and the surface temperature, we construct the length and timescales from $L_T$ (still unknown) and the prescribed convergence rate, and use the vertical normal stress at the base of the lithosphere as the stress scale. The latter choice is because differences between normal stresses, which contribute to deviatoric stresses, are indeed on that order. The dimensionless equation is thus:

\begin{equation}
0 \approx {{\nabla^2 T}} +{\dfrac{L_T^2\beta  g v_0}{\alpha_\textrm{th} C_p (T_s - T_0)}} {\sigma_{ij}}\left(\dot\varepsilon^\textrm{v}_{ij}+\dot\varepsilon^\textrm{vp}_{ij}\right)
\end{equation}

There is one dimensionless parameter at steady state, which we call heating or dissipation number. Moreover, this contains one variable quantity ($L_T$) the thermal measure of lithospheric thickness. We can estimate a steady-state $L_T$ by taking this dimensionless parameter to be of order unity, which amounts to equating our two (conductive and deformational) timescales. Deformational heating dominates until conduction is able to evacuate the heat to the surface at the same rate as it is produced, which requires the shallow geothermal gradient to steepen. The steady state should be reached on the timescale of the faster process, which is deformational heating. This leads to:

\begin{equation}
 L_T \approx \left({\dfrac{\alpha_\textrm{th} C_p (T_s - T_0)}{\beta  g v_0}} \right)^{1/2}
\end{equation}

For orders of magnitude we take: ${(T_S - T_0 )\approx 10^3\; \mathrm{K}, v_0\approx 10^{-9}\; \mathrm{m s^{-1}}}$, ${\alpha_\mathrm{th}\approx 10^{-6}\; \mathrm{m^2 s^{-1}}}$,
${\beta\approx 1}$,
$g\approx 10\; \mathrm{ms^{-2}}$,
$C_\mathrm{P} \sim 10^3 \;\mathrm{J kg^{-1} K^{-1}}$, which gives $ L_T\approx 10^{4}$ m. This seems small, but underlines the tremendous power developed during deformational heating and relative inefficiency of conductive heat removal. The deformed materials have become very hot quite close to the surface. In geological reality, near to the surface, hydrothermal systems can exploit the permeability offered by deformed rocks and enhance the evacuation of heat compared with pure conduction, which may modify this estimate. Nevertheless, the key takeaway from this analysis is that we might expect our system to be approaching a steady thermal state after roughly a few million years, at which point we might expect the thermal lithospheric thickness $L_T$ to be $\sim{10}$ km.

\subsection{Transient Evolution of Strain and Temperature}\label{stress_matching}
During the first 6 myrs, the convergence of 20 km/myrs has progressively produced a topographic ridge in the Caribbean plate and a parallel trough on the Atlantic side. The amount of vertical offset between ridge and trough that accumulates in our model system is on the order of 10 per cent of the convergence and therefore high. Assuming an initially uniform ocean of 5 km thick, the top of this ridge would be high enough to rise above sea level when the vertical offset between it and the upper surface of the plates where they are most downwarped, is approximately greater than 10 km. This is indeed the case after 5 myrs, which suggests that we might interpret the broad high on which the Lesser Antilles volcanic arc is built (Figure \ref{momoh_map}) as fundamentally due to buckling. The trough on the Atlantic side would be expected to fill up with sediment, which would attenuate the actual topographic offset with respect to that observed in the model.
\begin{figure}
\centering{\includegraphics[width=14cm]{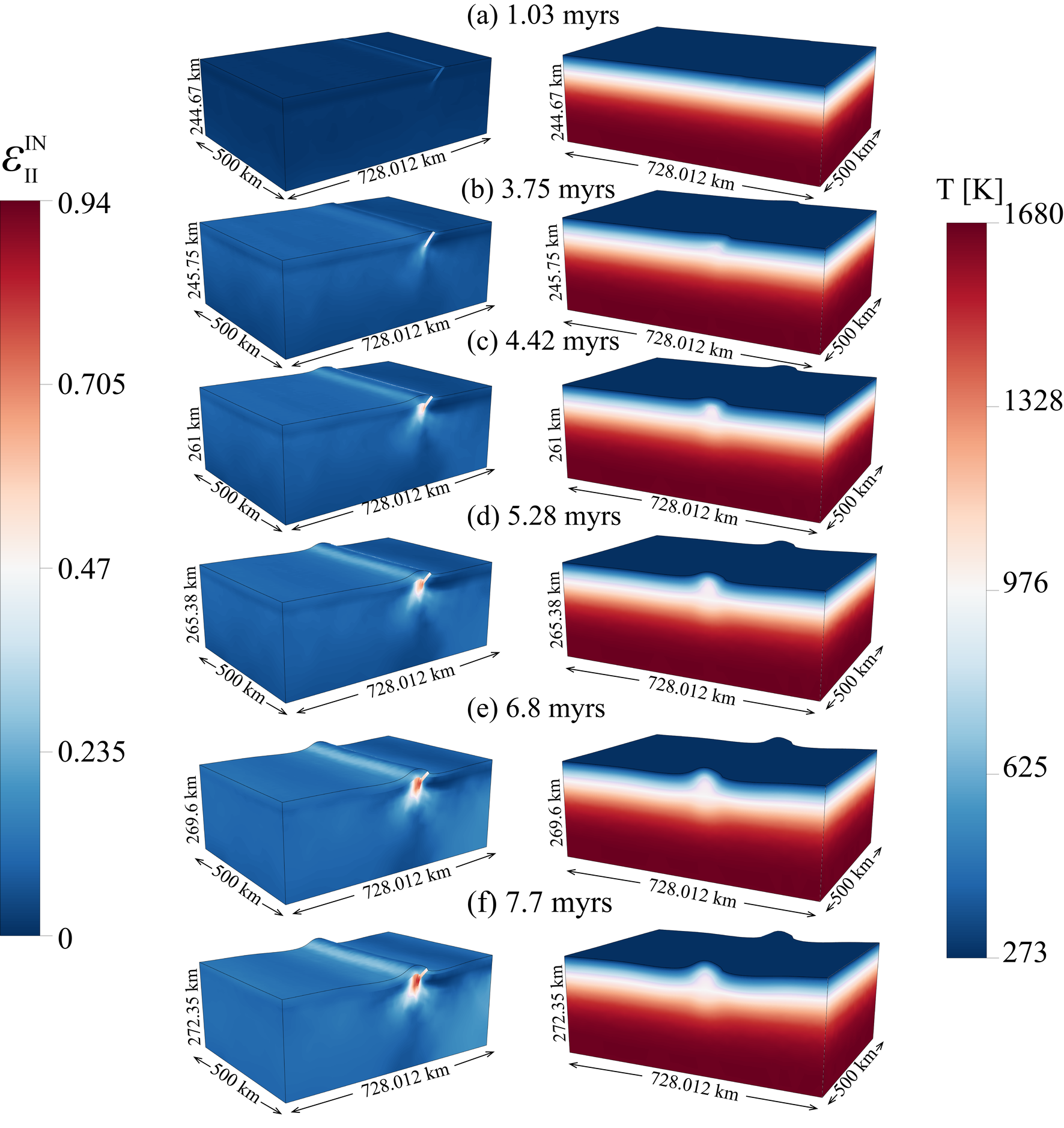}}
\caption{Evolution of inelastic deviatoric strain history (left panel) and temperature (right panel).}
\label{time_steps_evolution_of_EII_T}
\end{figure}

The first order response of the model is thus buckling, and hence the zone beneath the apex of the upward lithospheric fold is subject to intense irreversible deviatoric strain (see left panels in Figure \ref{time_steps_evolution_of_EII_T}). Given the physical framework we have laid out, we expect a temperature increase to occur there. While buckling is not a surprise, the heating effect turns out to be substantial. After $\sim$ 6 myrs of deformation, a thermal anomaly several tens of km wide with a central $\Delta T \sim$ 150-200 K (see right panels in Figure \ref{time_steps_evolution_of_EII_T}) has arisen. Consider the developing geometry of the zone of irreversible deformation and therefore heating. In the first 1-2 myrs, an inclined lithospheric plastic deformation band (nucleated on the initial weak zone) is prominent (Figure \ref{time_steps_evolution_of_EII_T}).  By 4.42 myrs, a second conjugate deformation band is visible on the other side of the topographic high, which intersects the first band beneath it, forming a v-shaped or an x-shaped feature (Figure \ref{time_steps_evolution_of_EII_T}c). This intersection of plastic deformation bands, which drive localisation in different directions, is the key feature that anchors the strong heating zone right beneath the high topography. Nevertheless, the plastic band on the Atlantic side of the ridge remains more pronounced, as might be expected given the initial weak zone. By 6 myrs, it extends into the asthenospheric mantle as an inclined zone of ductile shear, which is suggestive of subduction of Atlantic material. Buckling has led to downward flexure of the Atlantic plate. Although this basic situation suggests the correct dip direction of subduction, i.e., Atlantic under-thrusting Caribbean to the west, after several million years, there is no clearly developed cold slab.

\begin{figure}
\centering{\includegraphics[width=14cm]{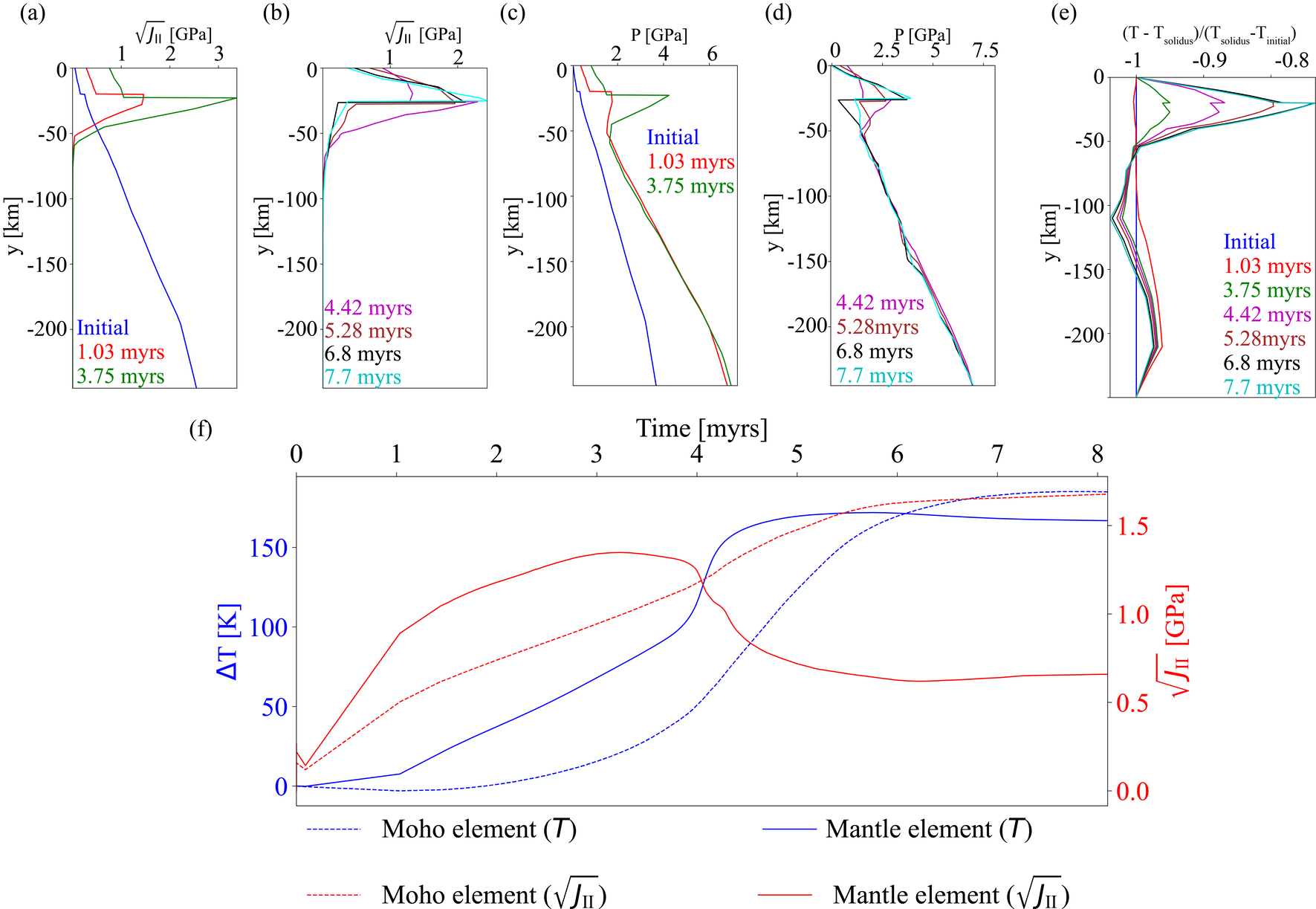}}
\caption{1-D plots extracted from the center of the peak topography for the evolution of: (a) Deviatoric stress invariant for a time window of 0 and 3.75 million years, (b) deviatoric stress invariant for a time window of 4.42 to 7.7 million years, (c) pressure for 0 to 3.75 million years interval, (d) pressure from 4.42 to 7.7 million years interval (e) normalised temperature. (f) Time evolution of temperature change from the initial temperature and deviatoric stress invariant evolution for an element extracted from the Moho beneath the elevated topography (broken lines) and within the mantle (${\sim}$46 km depth beneath the elevated topography (full line). Both the temperature and deviatoric stress invariant assumed steady state from 4 to 5 million years, accompanied by a rapid temperature rise for the same time interval. The deviatoric stress invariant shows a peak between 3 million years and drops from around 4 million years, where a rapid temperature rise is observed within the mantle element, while the Moho element shows a rising deviatoric stress invariant that flattens where the temperature change reaches a plateau.}
\label{time_steps_evolution_of_JII_P_DT}
\end{figure}

To further illustrate transient development during the first 5 to 7 myrs of convergence in the zone beneath the developing arc topography, Figures \ref{time_steps_evolution_of_JII_P_DT}a-e show vertical profiles of the deviatoric and normal stress invariants, i.e., $\sqrt{J_{II}}$ and pressure, as well as the temperature change relative to the initial geotherm. All show a maximum at or a bit below the Moho depth. The maximum $\sqrt{J_{II}}$ is a brittle-ductile transition. But the shape of the $\sqrt{J_{II}}$ curve in the $\sim$ 20-25 km just below the maximum $\sqrt{J_{II}}$, reflects the fact that differences between the normal stresses are driving both ductile creep and plasticity. The non-isochoric part of the deformation is being attributed here to plasticity whereas dislocation creep conserves volume. As intensive parameters evolve in a given rock volume, the rheologic response will change, for example as temperature increases, further inelastic deformation is accommodated by dislocation creep. If the latter becomes dominant, we could then say that sub-volume belongs to the asthenosphere, even if it started out in the lithosphere. However, the spatial transition from a predominantly lithospheric to predominantly asthenospheric response in this complex deformation zone occurs over several tens of kilometres of the uppermost mantle, within which the maximum $\sqrt{J_{II}}$ remains the relevant scale for deviatoric stress (Figure \ref{time_steps_evolution_of_JII_P_DT}a). We will refer to this interval as the Lithosphere-Asthenosphere Transition (LAT), whose thickness is an unknown a priori. In the sub-arc LAT, deviatoric stress level is 1 GPa, which, combined with strain rates (${\sim}$ 10\textsuperscript{-14} s\textsuperscript{-1}) implies a heating rate of $\sim$ 10\textsuperscript{-5} W m\textsuperscript{-3}. Sustained over a few million years, this leads to the observed substantial temperature increase. The thermal anomaly due to deformational heating grows to eventually affect the uppermost $\sim$ 100 km of the Caribbean plate with a maximum situated at roughly the Moho depth (Figure \ref{time_steps_evolution_of_JII_P_DT}e).

Figure \ref{time_steps_evolution_of_JII_P_DT}f shows the transient behavior of two elements selected for illustration, one close to the Moho and the other in the upper mantle at ${\sim}$ 50 km depth. The thermal anomaly associated with both these elements initially increases and then flattens to a steady value of $\sim$ 150-200 K, at $\sim$5 myrs for the mantle element and $\sim$7 myrs for the Moho element. The steepest increase occurs at 4-5 myrs. At steady state, conduction can evacuate the heat produced. There is an interesting contrast between the mantle and Moho elements in terms of stresses. Whereas for the Moho element $\sqrt{J_{II}}$ increases and then flattens off rather like $\Delta T$; for the mantle element, it reaches a maximum value at $\sim$4 myrs and then falls off before flattening. The reason for these maxima is that as heating occurs ductile creep and plasticity grow in importance and are able to reduce the differences between the normal stresses which contribute to both $\sqrt{J_{II}}$ and Pressure.

\subsection{Structure of the convergence at Thermal Steady State}
\label{quasisteadystate}
Figure \ref{time_steps_evolution_of_JII_P_DT}f points to a roughly steady thermal state having been reached by $\sim$ 5-6 myrs whereby conduction to the surface approximately balances deformational heating. Results corresponding to a simulation time of $\sim$ 8 myrs are shown in Figure \ref{momoh_output_model}. We show deformation (Figure \ref{momoh_output_model}a-c), distortional stress per unit volume (Figure \ref{momoh_output_model}d) temperature change from initial temperature (Figure \ref{momoh_output_model}e) and accumulated temperature (Figure \ref{momoh_output_model}f) within the model. We briefly describe the potential arc structure to provide a basis for the discussion in Section 4 which addresses our results in the light of Lesser Antilles data.

The deviatoric strain rate block (Figure \ref{momoh_output_model}a) shows maximum values (${\sim}$10\textsuperscript{-14} s\textsuperscript{-1}) occur in the uppermost mantle beneath the fold apex. This zone also shows large accumulated deviatoric strain, implying that this is a stable feature because it is anchored at the intersection of the two plastic deformation bands (Figure \ref{momoh_output_model}b).  We can identify this position, where topographic and thermal highs coincide as the likely magmatic arc area (Figure \ref{momoh_output_model}a-e). The presence of volumetric strain is a functional definition of the lithosphere given our assumptions that volumetric deformation is influenced by dilatant plasticity. On both Atlantic and Caribbean sides of the model, away from the region of buckling, volumetric deformation is quite uniformly restricted to the uppermost 50 km. This can be taken as the far-field lithospheric thickness where deformation has not localised, given the initial geotherm and kinematic conditions. Beneath the arc, where deformation has localised, volumetric deformation extends to greater depth (70 km) and a large amount of deviatoric strain has accumulated (Figure \ref{momoh_output_model}b,c), i.e., the lithosphere bulges downwards where deformation is more intense. The uppermost mantle rocks belonging to the lithosphere and the lowermost crust are the vertical interval in which ${\sqrt{J_{II}}}$ has its highest values (Figure \ref{momoh_output_model}d) which combined with the strain rate distribution results in the thermal structure (Figure \ref{momoh_output_model}e,f). (Figure \ref{momoh_output_model}a) also shows a ductile shear zone dipping beneath the arc but which is without volumetric deformation. We also note a tensional deformation feature occurring at the apex of the arc connecting the hot zone below to the surface.  

\subsection{Partial Melting and Volcanic Arc Position}
\label{partialmeltingimplications}
A quantitative assessment of partial melting within the above structure requires knowledge of the absolute value of temperature and not simply the increase in temperature, as well as the appropriate value of the solidus, crucially dependent on the poorly known amount of water or other volatiles present. For simplicity in these simulations we used an anhydrous solidus \cite{hirschmann2000mantle,katz2003new}. The initial geotherm is not well enough constrained to be confident of absolute temperature values. It is highly plausible that some volatiles (H\textsubscript{2}O and CO\textsubscript{2}) would be present in the source which greatly lowers the solidus (of both crustal and mantle lithologies). Nevertheless, our results hint strongly at where a magmatic arc is likely to form, if volatiles are present or start to be supplied from a subducting slab into this thermal structure. Beneath this arc region in the mantle, small amounts of partial melt (0.62 to 3.2 per cent) of anhydrous peridotite were observed in our simulations (Figure \ref{momoh_output_model_partial_melt}).  The depth interval of the mantle most affected by deformational heating is 20 to 50 km. This interval also experiences decompression between about 4 and 6 myrs which would assist melting. However, melting at these shallower depths might only occur in the case of a hydrous mantle solidus. The volatiles necessary for melting could either be initially present or be introduced into this pre-heated zones once de-volatilization reactions in subducting material commence. 

\begin{figure}
\centering{\includegraphics[width=15cm]{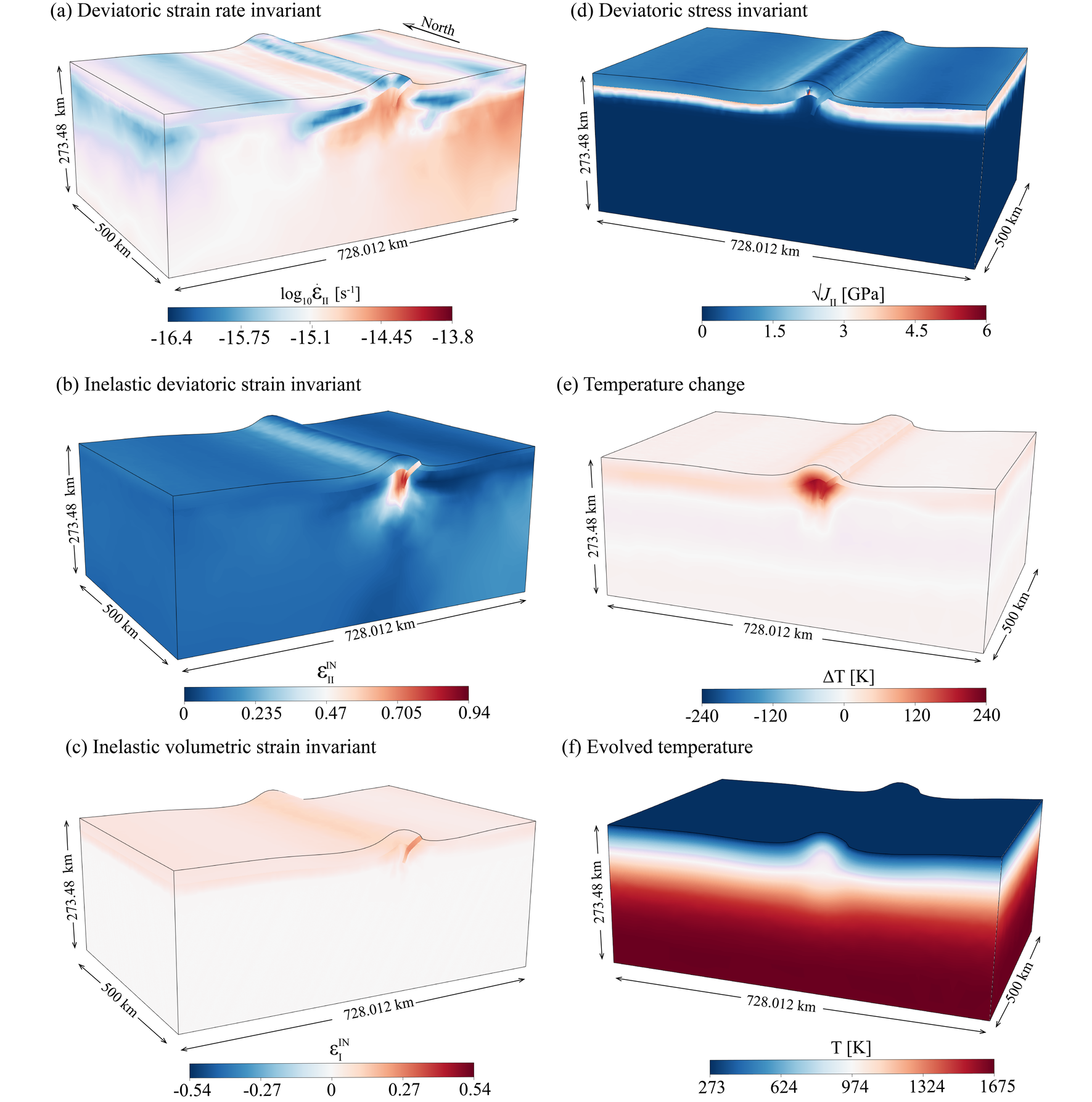}}
\caption{Deformation state (a to c), temperature state (d), and stress state (e). Simulation was run for 7.7 million years at a compression rate of 2 cm/year, held constant.}
\label{momoh_output_model}
\end{figure}

In terms of horizontal position, our results suggest the magmatic arc would be expected to initiate/develop on the high topography, which is fundamentally controlled by the wavelength of lithospheric flexure. The flexural length of the Caribbean lithospheric plate can be estimated using \cite{watts2001isostasy}:

\begin{equation}
\lambda =\left({\dfrac{E L_T^3}{3g\Delta\rho (1-\nu^2)}} \right)^{1/4}
\end{equation}

as long as $L_T$ is known. We take the far-field lithospheric thickness observed in our results, ${L_T\approx}$ 50 km. Using the following orders of magnitude:
${E\approx 10^{11}}$\;Pa, $g\approx 10\; \mathrm{m\;s^{-2}}$, ${\Delta\rho \approx 10^{3} \;\mathrm{kg\;m^{-3}}}$, and $\nu=$0.25,
leads to $\lambda\approx 10^{5.25} \approx$ 180 km. We thus expect this flexural length to correlate with the distance between the topographic maximum (i.e., the volcanic arc) and the interface between the plates. Note the latter does not correspond to the bathymetric trench position, but is closer to the arc and buried beneath thick sediments. In terms of depth, the maximum temperature increase occurred at and around the Moho. Because the Caribbean crust is thick ($\approx 20$km), the initial Moho temperature is relatively high, and as the lowermost crustal rocks have been strongly heated, this hints at partial melting of the lower crust; although we did not include a crustal solidus for now. The temperature rise beneath the high topography is of the order of 226 K (Figure \ref{momoh_output_model}e). To actually observe partial melting of the crust, a solidus depressed by volatiles would need to be used.

\begin{figure}
\centering{\includegraphics[width=15cm]{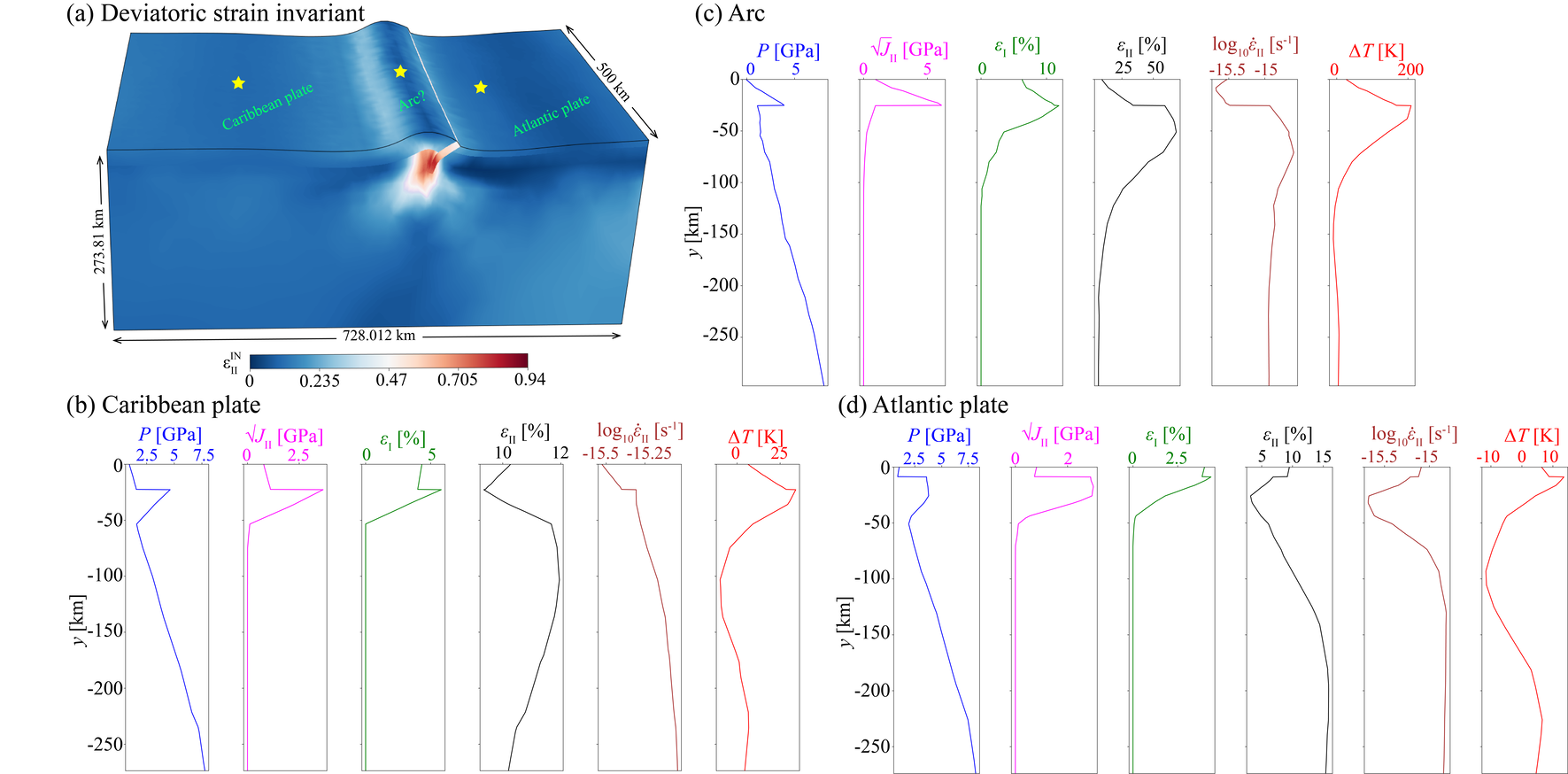}}
\caption{1-D extracts of stress state, deformation state and temperature effects. (a) Deviatoric strain invariant showing locations of paths for quantities shown for (b) Caribbean plate, (c) arc area and (d) Atlantic plate.}
\label{profile_extracts}
\end{figure}
\begin{figure}
\centering{\includegraphics[width=12cm]{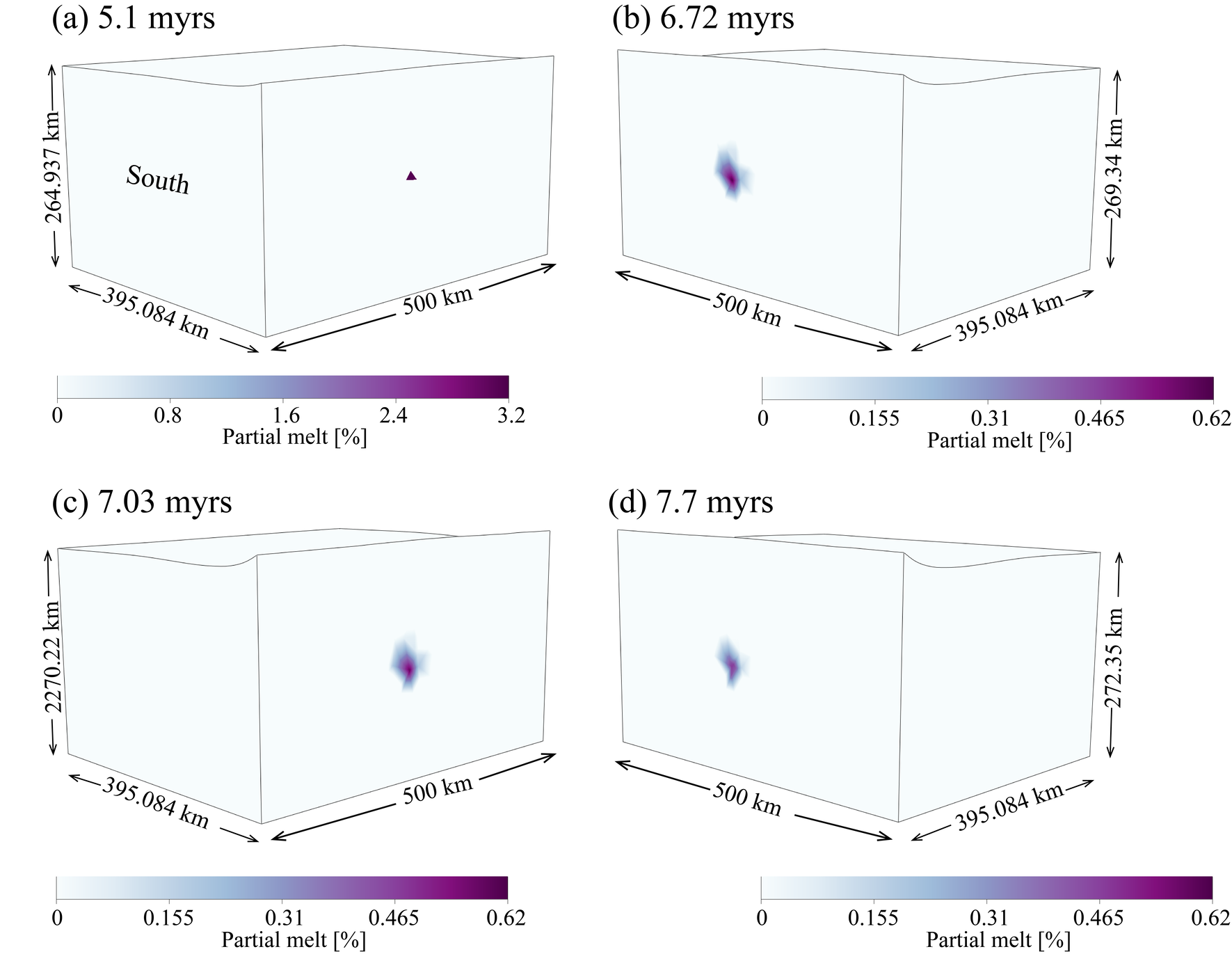}}
\caption{Partial melt within the mantle at different model times below the  Caribbean plate (left panel) and 46 km below the peak elevation (right panel). The vertical slice was taken in the Caribbean plate, and the elevated topography in the left and right panels, respectively.}
\label{momoh_output_model_partial_melt}
\end{figure}

\section{Discussion}
We now assess the outcome of our modelling for the transition from plate convergence to stable subduction in the light of data from the Lesser Antilles, highlighting the following points:
\begin{enumerate}
\item Timescale of subduction initiation
\item Lithosphere Asthenosphere Transition (LAT), deformational heating magnitude and thermal steady state
\item Magmatic arc position and topography due to compressional-deformation
\item Partial melting from an energetic standpoint
\end{enumerate}

\subsection {Timescale of subduction initiation (LAT)}
A key plank of our approach was not to presuppose the geometry of steady subduction but to begin from an undeformed state. We can probe the model as to what timescale a plausible subduction-like interplate geometry develops and what topographic and thermal structure exists by the end of this initiation phase. We found that a roughly steady thermal state was established in ${\sim}$5 myrs, which reflects the deformation timescale (Section \ref{timescaleevolution}). It is less easy to identify a deformational steady state. Although the timelapse between the onset of convergence and the onset of volcanism is not tightly constrained for the Lesser Antilles Arc, a period of a few million years is consistent with existing estimates (see \citeA{allen2019role}). Moreover, this value compares favourably with the more general observational timescale that emerges from the comprehensive review of subduction initiation events during the last 70 myrs \cite{lallemand2021subduction}. The criterion used by those authors to assess whether a given sequence actually achieved sustained subduction is the onset of volcanism. They did emphasise buckling but did not include deformational heating and assumed that volcanism must imply slab formation. It is not obvious that a deep slab can form and begin to release volatiles in the space of a few million years. We found that by ${\sim}$5 myrs, the buckling response of the plates and associated deformational heating produced a thermal anomaly possibly strong enough to initiate partial melting beneath a topographic ridge, already high enough to be emergent. The initiation of volcanism could therefore pre-date the development of a deep slab supplying volatiles, and the thermal anomaly pre-dates volcanism. After this initial phase it seems reasonable that down-going material and an associated corner flow will develop, as has been modelled in many numerical models of steady subduction which assume the slab to be present as a kinematic feature and gloss over the initiation phase. We do not observe the expected strong negative thermal anomaly beneath the arc, which could potentially drive steady buoyant convection. But our results could be consistent with negative buoyancy being generated by mineral (notably devolatilization) reactions that are to be expected, but this is a level of complexity that we did not address at this stage. We highlight below that the fundamental structure that emerges by ${\sim}$5-6 myrs, is to first order consistent with the currently observed Lesser Antilles, which is presumed to be in steady subduction. Deformational heating turns out to be remarkably strong, contributing in excess of 200 K to localised temperature change within the deformation zone. The thermal structure of steady subduction can thus be strongly influenced by the transient initiation phase.

\subsection {Lithosphere-Asthenosphere-Transition (LAT) beneath the Lesser Antilles Arc (LAA)}
Our approach allows lithosphere and asthenosphere to emerge from the initial state as local temperature, stress and deformation mechanisms evolve. As in Section \ref{modelling_and_results}, we take the zone undergoing volumetric deformation, as a working definition to identify the lithospheric domain in our results, i.e., that capable of brittle rupture, which we can compare with the envelope of tectonic seismicity. 

By the end of the initiation phase, as steady subduction takes over, the depth of onset of ductility has shallowed beneath the arc. The maximum $\sqrt{J_{II}}$ has localised around the Moho, which is a major rheological boundary under conditions of strong deformation (Figure \ref{time_steps_evolution_of_JII_P_DT}a), which would lead us to expect enhanced seismicity at around Moho depth beneath the volcanic arc. This is consistent with locations in the seismic catalogue of the French observatories in Martinique and Guadeloupe. For example, in Martinique, little to no crustal seismicity occurs in the depth interval $\sim{5-15}$ km whereas numerous events occur in the interval $\sim{20-40}$ km, with the Moho depth estimated at $\sim{25}$ km from receiver functions \cite{schlaphorst2018probing}.
While the onset of ductility has shallowed, the depth at which volumetric strain disappears has deepened beneath the arc (Figure \ref{momoh_output_model}c). Although the arc is hotter, tectonic seismicity should thus extend to greater depths beneath it, which is observed. The data show an asymmetric V-shaped envelope of tectonic seismicity whose lower tip is at $\sim$70 km depth, horizontally offset from the volcanic arc towards the trench $\sim{50-70}$km \cite{paulatto2017dehydration}, and identified with the deepest part of the subduction interface \cite{bie2020along}. The envelope of volumetric deformation from our results compares quite well with that of the observed tectonic seismicity (Figure \ref{model}). Our figure is constructed using the catalogue of the French observatories, but the  basic distribution is the same as in the above references which combined data from Ocean Bottom Seismometers (OBS) deployments with land stations to relocate seismicity. At depths ${\ge}$70 km beneath the arc, we find temperatures too high for lithospheric behavior and essentially all shear is by dislocation creep. There is of course a tongue of intermediate to deep seismicity dipping steeply down from the cusp of the cluster of tectonic events to $\sim${200 km}, which locates the Atlantic slab \cite{bie2020along}. Source mechanisms of such deeper seismicity are often ascribed to mineral transformations such as eclogitization and/or de-volatilisation rather than classic tectonic seismicity e.g., \cite{paulatto2017dehydration}. The precise rheology of down-going material undergoing mineral reactions is not easy to characterize. Large scale deformation may be predominantly ductile/asthenospheric whilst mineral reactions nevertheless drive seismicity. We do not thus equate all seismicity with rheologic lithosphere but the distinction is not easy to make.
Our model cannot explicitly address this. 

\subsection{Horizontal arc structure on reaching steady state}
The obvious steady deformational state is the classic picture of kinematically under-thrusting material that accommodates virtually all the convergence, and may be the best way to extrapolate from our results towards a more familiar steady subduction picture. The diffuse zone of ductile shear extending downwards into the asthenosphere (Figure 5a) hints at the kind of structure usually taken as a starting point for steady subduction models with a kinematic condition in the slab surface. The two approaches are mostly complementary. Continued subduction might then occur with little or no more topographic uplift. The stress beneath the arc appears to stabilise in our results (Figure \ref{time_steps_evolution_of_JII_P_DT}f), and we might infer that deformational heating has stabilised and is balanced by conduction to the surface. 

\begin{figure}
\centering{\includegraphics[width=15cm]{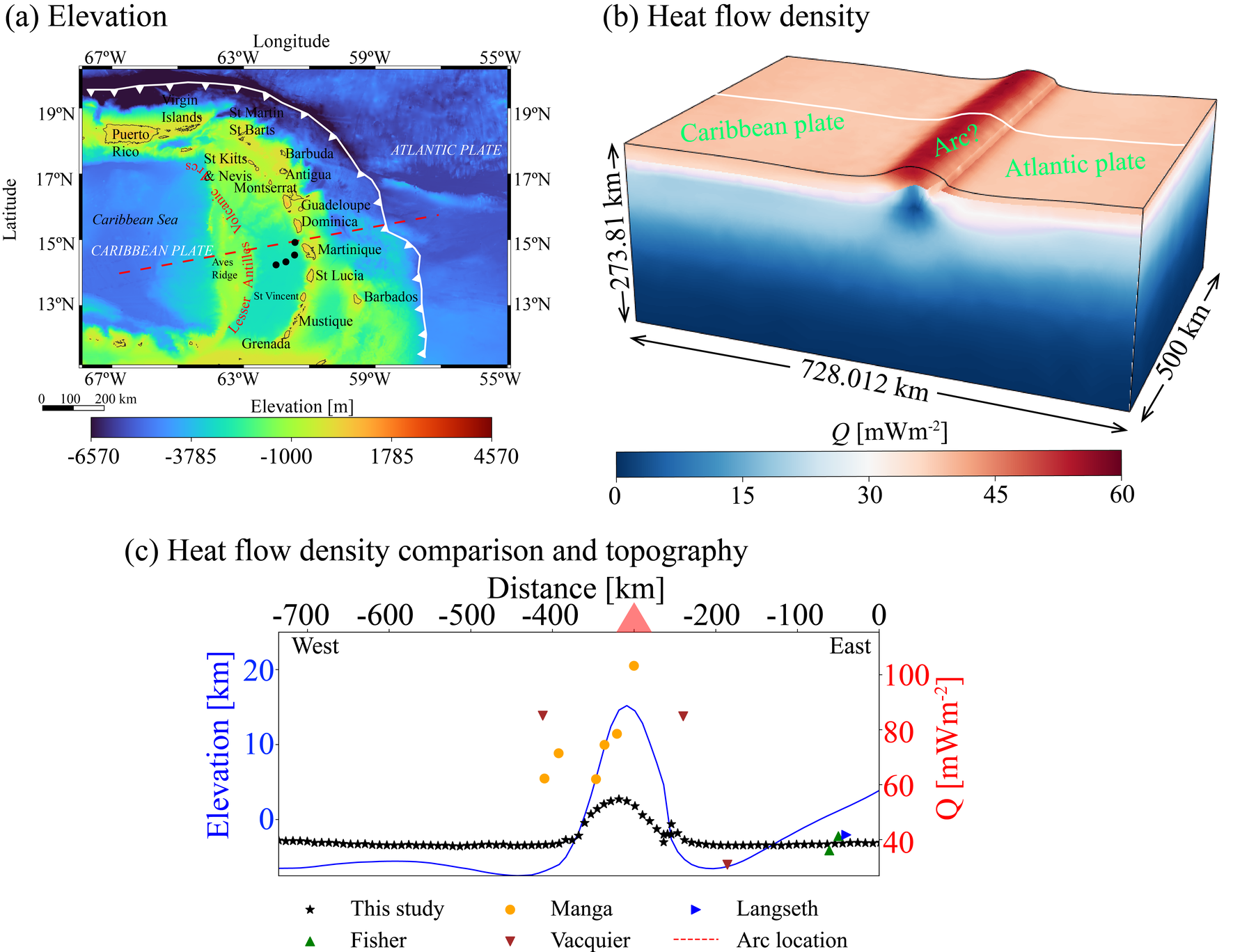}}
\caption{(a) Topographic map for the Lesser Antilles area showing the volcanic archipelago and the subduction trench (in white). The red dashed line outlines a profile for which heat flow transects were taken and black filled circles are actual heat flow measurement locations \cite{manga2012heat}. (b) Heat flow density model from our simulations with white line along which topography and surface heat flow density values are extracted and plotted in (c) to compare modelled heat flow density values and actual heat flow density values from the literature. (c) Heat flow density values shown in coloured symbols with corresponding axis on the right, and topography (blue line) with axis on the left. The red triangle at a distance of 300 km from the East is the location of the arc. Our heat flow density profile corresponds approximately to the red dashed line shown in (a).}
\label{heatflowcompare}
\end{figure}

\begin{figure}
\centering{\includegraphics[width=13cm]{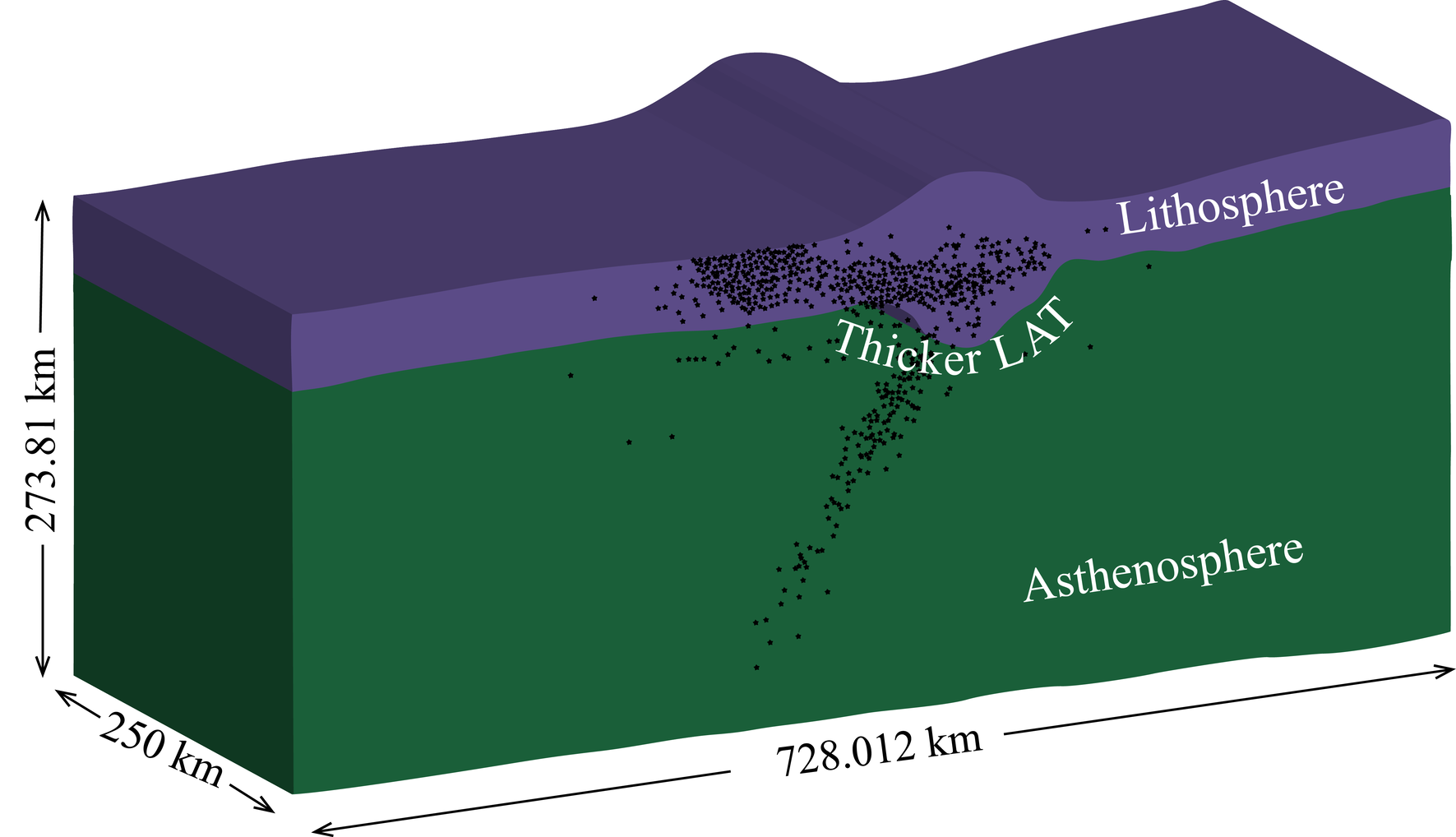}}
\caption{Model showing outline of the lithosphere and asthenosphere, and sketch of low-magnitude earthquakes from the Lesser Antilles. The inelastic volumetric strain, thermal anomaly and deviatoric strain have been used to determine the boundaries of the lithosphere. The thick edifice is characterised by a thicker lithosphere-asthenosphere transition zone.}
\label{model}
\end{figure}

From \citeA{kopp2011deep}, the horizontal separation between the volcanic arc and the subduction interface where the subducting Atlantic plate surface passes from contact with sediments to contact with the Caribbean oceanic crust is ${\sim}$ 170 km, which is consistent with our order of magnitude estimate (Section \ref{partialmeltingimplications}) based on buckling of a ${\sim}$50 km thick plate. \citeA{kopp2011deep} also show the sediment-covered, down-warped inter-plate contact at a depth of ${\sim}$15 km. Remarkably, their seismic velocity structure is horizontally quite uniform, implying that the Caribbean crust near the subduction interface is similar to that beneath the arc. The other observational test of this horizontal scale of subduction is the width of the band of tectonic seismicity. The overall width is 200-250 km of which about 170 km lies east of the volcanic arc \cite{gutscher2013wide,paulatto2017dehydration,bie2020along}, again consistent with the width of the buckled zone either side of the inter-plate contact, as evoked in models of bending of incoming plates. The Atlantic plate is the more rigid of the two, due to its lower crustal thickness. As already observed (Figure \ref{heatflowcompare}c), the Lesser Antilles islands are the emergent crest of a much broader bathymetric high, which we identify with the upward fold of Caribbean lithosphere observed in the simulations, and on top of which the Lesser Antilles volcanoes have been built. The distance between the inter-plate contact and the arc is a clearer measure of mechanical horizontal length scale of subduction rather than the arc-trench distance, because the latter strongly reflects the sedimentary infill of the basin formed by downward plate flexure. Indeed, the abrupt increase, from North to South, of the arc-trench distance at the latitude of Martinique, which is a peculiarity of the LAA, reflects mainly the greater sedimentary flux in the south coming from large South American rivers. By contrast the mechanical subduction interface as revealed by seismicity is at a relatively constant distance from the arc \cite{gutscher2013wide}.



Heat flow measurements in the Lesser Antilles Arc have been undertaken by a number of workers \cite{manga2012heat}. Figures \ref{heatflowcompare}b and \ref{heatflowcompare}c compare our numerical results with a compilation of the available measurements. Variations in technique and other factors which may influence measurements and errors are present; the data are broadly consistent where they overlap, and a general picture emerges. The data show an interesting variation in the direction normal to the arc. 
Heading from the trench towards the arc heat flow is ${\sim}$30-40 mWm\textsuperscript{-2}, increasing steeply to ${\sim}$70-80 mW m\textsuperscript{-2} over a horizontal distance of about the last 100 km leading up to the volcanic arc, with one somewhat isolated peak value of ${\sim}$100 mW m\textsuperscript{-2}. For comparison, we show the heat flow pattern that we obtained from our simulations at ${\sim}$ 7.7 myrs. The major positive thermal anomaly is predicted to be at the arc, i.e., coinciding with the topographic maximum. Quantitatively, our results are of the right order of magnitude, but only get within approximately a factor of two of the observations. The overall picture of a roughly Gaussian horizontal profile in the heat flow centred on and correlated with the high topography of the arc, is qualitatively consistent with observations (Figure \ref{heatflowcompare}c). The message contained in the data is that such relatively short wavelength variations must reflect quite a shallow thermal structure.

\subsection{Arc establishment, High heat flow with small partial melt as a consequence of deformation}
The analysis of \citeA{perrin2016reconciling} raised the interesting point that the thermal structure developed in the steady corner flow model is insufficient to sustain the level of heat flow observed, i.e., the entrained mantle wedge does not supply enough heat. They showed that petrologic constraints suggest a shallow hot zone with mantle partial melting below the LAA at pressures of ${\sim}$1-2 GPa which could not be explained by the corner flow model alone. These melting pressure values are broadly consistent with the pressures we calculated for the uppermost ${\sim}$ 20-40 km of the sub-arc mantle (Figure \ref{profile_extracts}c). The temperatures attained in our model after ${\sim}$6-7 mys are still less than those required by the data of \citeA{perrin2016reconciling}, but the location and shape of the additional thermal anomaly produced by deformational heating looks correctly situated compared with the conclusions of \citeA{perrin2016reconciling}. The small amount of partial melting in our model occurs at ${\sim}$100 km, beneath the Moho - clearly in the asthenosphere, and deeper than the above estimates - where pressures are more like ${\sim}$3-4 GPa. This discrepancy could partly be because our initial geotherm was too cold but also because we used the anhydrous solidus. Given the large uncertainty in solidus we cannot address this important question quantitatively as yet. One difference of our simulations regarding precise comparisons with the Lesser Antilles data is that the true convergence is oblique. A large-scale component of along-strike shear, due to oblique convergence, might accentuate localisation and quantitatively affect heating rates. { {The obliquely subducting fracture zones in the Atlantic plate \cite{feuillet2002arc} are likely to provoke additional localisation of deformation and contain more volatiles in serpentinite \cite{davy2020wide}}}. The takeaway is that the thermal structure we have found predates volcanism; it is not caused by it. Once melting and melt transport initiate, subsequent advection of heat will presumably modify or even accentuate the thermal structure along the lines proposed by \cite{jones2018thermal} who sought an explanation for the heat flow in excess of that predicted by the steady corner flow model.

Based on our results, partial melting in the mantle is most likely to occur not far beneath the Moho, say in a depth interval of $\sim$ 20 - 50 km. The lower Caribbean crust is also anomalously hot, probably ductile and may also be close to or even above its solidus depending on composition, but we did not yet explore this in detail. Partial melting of crustal rocks would be expected in the interval just above the Moho, in this case at depths of $\sim$ 20 km. Based on \citeA{sapegina2025basalt} on the melting of basaltic rocks as a guide, we can estimate that hydrated basaltic rocks at the pressures and temperatures we find at the Moho, i.e., $\sim$ 1-2 GPa and ${\sim}$1023 K lower crustal rocks might contain about 20 per cent of rhyolitic melt. As is well known, the eruptive fluxes are predominantly silicic in the more productive central part of the arc, including the Saint Lucia and Guadeloupe arc segment, for example \cite{wadge1994lesser}. Our results thus hint at independent melting of lower crust and upper mantle. Given the uncertainties, we cannot yet assess magma fluxes of these two magma types quantitatively. However, this situation seems more consistent with a dearth of intermediate melt compositions in arc volcanics except by mixing as pointed out by  \citeA{reubi2009dearth} to be a more general feature of arc volcanics. Recent eruptives from Mont Pelée contain rhyolitic glass as well as mafic enclaves \cite{pichavant2002physical}.

{{During the last two decades, imaging of the Lesser Antilles has been greatly advanced by a number of projects, often with focus on the paths that volatiles released from a deep slab might follow and drive flux melting assuming a steady-state subduction configuration. Our work cannot directly address this issue as we focus on the establishment of the thermal structure (presumably mostly) prior to the onset of volatile release, but into which volatiles will be released once dehydration reactions eventually initiate. Nevertheless, the recent tomographic images \cite{paulatto2017dehydration,hicks2023slab,bie2022imaging,schlaphorst2021variation} clearly show interesting along arc variability. Melt inclusion studies also found that LAA melts contain variable volatile contents up to quite high values \cite{cooper2020variable}, indicating a variably hydrated source. Those results emphasised the impact of spatially variable volatile supply because of variable serpentinisation, notably linked to fracture zones in the Atlantic plate \cite{schlaphorst2016water}. The general 3D character of the observed structures is to be expected within the physical framework we are putting forward to study localisation of deformation and associated energy release. But in order for precise comparisons to be realistic, inclusion of more detailed 3D initial conditions than has been attempted here would be required. The conventional 2D steady models invoking a deep kinematic slab are not able to explain the data on thermal structure and partial melting \cite{perrin2016reconciling}.}}

The Galapagos hotspot has been invoked as a source of heat for a hot asthenosphere migrating eastwards towards the Caribbean and altering the pre-existing topographic expression \cite{chen2021caribbean}. Plume-induced subduction initiation has been studied numerically where a hot plume was invoked to break through the oceanic lithosphere, assuming the plate is sufficiently weak \cite{baes20163,baes2020plume}. Our results lead us to propose an alternative hypothesis, namely that a prominent source of heat is also inelastic creep and plastic deformation that occurs due to buckling often present during subduction initiation \cite{lallemand2021subduction}. We show that atop the strong creep deformation band in the mantle is a buckled surface bounded by two crustal plastic deformation bands at whose intersection a large thermal anomaly arises from heat dissipation during irreversible creep deformation in the ductile mantle and plastic deformation in the brittle crust (Figure \ref{momoh_output_model} and Figure \ref{momoh_output_model_partial_melt}). We find a temperature rise in excess of 226 K within 8 million years.

\section{Conclusion}
\begin{enumerate}
\item While a common argument for steady-state subduction (i.e., a deep slab) in the literature is the formation of volcanic arcs, we have shown here that deformational heating can generate high heat flux without necessarily melting rocks. We have also shown that an arc-like geometry forms and that this heating may be enough to generate partial melt in both sub-arc mantle and initiate volcanism within only a few million years, i.e., likely prior to steady subduction. 
\item When there is long-term compression, the heat generated due to inelastic plastic strains can cause an increase of 200 K within a few million years. The broad deformed sub-arc zone sees deviatoric stress and pressure drop after a rapid heating episode, which stabilises heat production. Such a pressure drop lowers the melting point, compounding the effect of volatiles. 
\item While common decompression melting argues in favour of rocks brought from high-pressure zones to lower-pressure zones, our results demonstrate that rocks can also partially melt when deformed locally. 
\item The evolution of the thermal field prior to the development of a corner-flow temperature distribution is a key highlight of our modelling.
\item Inelastic volumetric strain (defined for a brittle rheology here) offers us a proxy to define the lithosphere (undergoing elastic and plastic behaviour) and the asthenosphere (undergoing ductile creep). Beneath the developing arc, lithospheric thicknesses range between 50 and 100 km, and 50 km in the Atlantic and Caribbean plates.
\end{enumerate}

\section*{Acknowledgments}

This work received support from the Institut Physique du Globe de Paris (IPGP), INTERREG V Cara\"{i}bes program, the European Regional Development Fund (FEDER through the PREST project. EM benefited immensely from the AXA Research Fund through the prestigious AXA Postdoctoral Research Fellowship, and was hosted by G\'{e}oscience Environment Toulouse/Observatoire Midi-Pyr\'{e}n\'{e}es. Funding support and hosting are gratefully acknowledged. HSB appreciates the European Research Council grant PERSISMO (grant number 865411) for partially supporting this work. ST is very grateful for many stimulating meetings and discussions with members of the VOILA science team and colleagues linked with the French Antilles Observatories and the University of the West Indies.  This is IPGP contribution number: ....

\clearpage

\clearpage
\renewcommand*{\bibfont}{\scriptsize}
\printbibliography
\end{document}